\documentstyle[12pt,psfig,draft,lscape,array,amssymb]{nature-ejb}

\def\simlt{\mathrel{\hbox{\rlap{\hbox{\lower4pt\hbox{$\sim$}}}\hbox{$<$}}}}
\def\simgt{\mathrel{\hbox{\rlap{\hbox{\lower4pt\hbox{$\sim$}}}\hbox{$>$}}}}

\def\ale{\mathrel{\hbox{\rlap{\hbox{\lower4pt\hbox{$\sim$}}}\hbox{$<$}}}}
\def\age{\mathrel{\hbox{\rlap{\hbox{\lower4pt\hbox{$\sim$}}}\hbox{$>$}}}}

\newcommand{\sgras}{Sgr\,A$^\star$\,}

\begin{document}

\title{\bf A size of $\sim$1 AU for the radio source \sgras at the centre of the Milky Way}

\author{
Zhi-Qiang Shen\affiliation[1]{\scriptsize Shanghai Astronomical
Observatory, 80 Nandan Road, Shanghai 200030, China}, K. Y.
Lo\affiliation[2]{\scriptsize National Radio Astronomy
Observatory, 520 Edgemont Road, Charlottesville, VA 22903, USA},
M.-C. Liang\affiliation[3] {\scriptsize Division of Geological and
Planetary Sciences, California Institute of Technology, Pasadena,
CA 91125, USA}, Paul T. P. Ho\affiliation[4]{\scriptsize
Harvard-Smithsonian CfA, 60 Garden Street, Cambridge, MA 02138,
USA}$^,$\affiliation[5]{\scriptsize Institute of Astronomy \&
Astrophysics, Academia Sinica, PO Box 23-141, Taipei 106, Taiwan,
China},J.-H. Zhao\affiliationmark[4]}
\date{\today}{}
\headertitle{} \mainauthor{Shen et al.}

\summary{Although it is widely accepted that most galaxies have
supermassive black holes (SMBHs) at their centers$^{1-3}$,
concrete proof has proved elusive. Sagittarius A$^\star$
(\sgras)$^4$, an extremely compact radio source at the center of
our Galaxy, is the best candidate for proof$^{5-7}$, because it is
the closest. Previous Very Long Baseline Interferometry (VLBI)
observations (at 7mm) have detected that \sgras\ is ~2
astronomical unit (AU) in size$^8$, but this is still larger than
the "shadow" (a remarkably dim inner region encircled by a bright
ring) arising from general relativistic effects near the event
horizon$^9$. Moreover, the measured size is wavelength
dependent$^{10}$. Here we report a radio image of \sgras\ at a
wavelength of 3.5mm, demonstrating that its size is $\sim$1 AU.
When combined with the lower limit on its mass$^{11}$, the lower
limit on the mass density is 6.5$\times$10$^{21}$ Msun pc$^{-3}$,
which provides the most stringent evidence to date that \sgras\ is
an SMBH. The power-law relationship between wavelength and
intrinsic size (size $\propto$ wavelength$^{1.09}$), explicitly
rules out explanations other than those emission models with
stratified structure, which predict a smaller emitting region
observed at a shorter radio wavelength. }

\maketitle


\end{document}